\documentclass[11pt,a4paper]{article}
\usepackage{bm}
\usepackage{graphicx}
\usepackage{graphics}
\usepackage{float}
\usepackage{amsmath}
\usepackage{amssymb}
\usepackage{amsfonts}
\usepackage{latexsym}
\usepackage{epsfig}
\usepackage{dcolumn}
\usepackage{url}
\usepackage[francais,english]{babel}
\usepackage{lmodern} % math, R_m, ss, tt
\usepackage[T1]{fontenc}
\usepackage{pdfsync}% enable tex source and pdf output synchronicity
\usepackage[pdftex,bookmarks,colorlinks,breaklinks]{hyperref}
\hypersetup{linkcolor=black,
citecolor=blue,filecolor=red}
\newcommand{\Hcal}{{\mathcal H}}

\newcommand{\Lcal}{{\mathcal L}}
\newcommand{\Scal}{{\mathcal S}}

\newcommand{\Cc}{{\mathbb C}}

\newcommand{\half}{{\textstyle \frac{1}{2}}}

\usepackage{epstopdf}% to include eps graphics files with pdfLaTeX
\title{ %\textsc{
Some aspects of defect theory in spacetime}
\vspace{0.5cm}  

\author{\textsf{M. Kleman}\footnote{\textsf{kleman@ipgp.fr}} 
 \vspace {15pt}\\
Institut de Physique du Globe de Paris $-$ 
 Sorbonne Paris Cité\\
 1, rue Jussieu - Paris cedex 05, France}
 \date{} % Delete this line to display the current date
  
\begin{document}
\maketitle
%\newpage
\scriptsize
\tableofcontents
\newpage
\normalsize

\begin{abstract}
The topological theory and the Volterra process are key tools for the classification of defects in Condensed Mater Physics. 
We employ the same methods to classify the 2D defects of a 4D maximally symmetric spacetime. These \textit{cosmic forms}, which are continuous, fall into three classes: i)- $m$-forms, akin to 3D space disclinations, analogous to Kibble's cosmic strings; ii)- $t$-forms, related to hyperbolic rotations; iii)- $r$-forms, never considered so far, related to null rotations. A detailed account of their metrics is presented. There are \textit{wedge} forms, whose singularities occupy a 2D world sheet, and \textit{twist} forms, whose singularities occupy a 3D world shell. $m$-forms are {compatible} with the cosmological principle of \textit{space} homogeneity and isotropy, $t$- and $r$-forms demand  \textit{spacetime} homogeneity. $t$- and $r$-forms are typical of a vacuum obeying the perfect cosmological principle in a de Sitter spacetime. 
Cosmic forms may assemble into networks generating vanishing curvature.
\end{abstract}

\section{Introduction} \label{int}

This article aims at the application to Cosmology of the concepts of the Condensed Matter Physics (CMP) theory of defects, including the Volterra process (VP) for line defects  \cite{friedeldisloc}, the topological theory (TT) for quantized defects of any dimensionality (point defects, walls and solitons, configurations) \cite{toulouse76,kleman78}, and the continuous theory (CT) of line defect densities \cite{kroner,klfr08}. 
  
  The TT is not absent from to-day Cosmology, (cf. Kibble's \textit{cosmic strings} \cite{kibble76}, reviews in \cite{vilenkin94,hindmarsh95}); the analogies between quantized defects in liquid crystals and cosmic strings have been stressed and have inspired 'cosmology in the laboratory' experiments \cite{chuang91}. However its use has been limited; cosmic strings do not cover all the possible line defects allowed by the full theory. The VP has permitted to recognize cosmic string types not present in Kibble's theory \cite{hiscock85,tod94,puntigam97}.
 
The CT has been developped for spacetimes endowed with torsion, see a review of the Einstein-Cartan Theory (ECT) in \cite{hehl}. In \cite{ruggiero03} it is shown that a spacetime with curvature and torsion can be considered as a state of a four-dimensional continuum containing defects, in analogy with the CMP continuous theory of \cite{kroner}. In \cite{kleinert05} it is claimed that gravitation stems from a distribution of continuous defects. 
 
 But a full use of the TT is new; among other results, it yields a classification of \textit{}{cosmic defects} of various dimensions. The \textit{line} defects (\textit{cosmic forms}) discovered in conjunction with the VP (all are continuous defects) suggest new physical features of the spacetime. This is the content of this article.
\\ \vspace{-10pt}

 Let us first give a short reminder of the theory of defects (for a detailed overview see \cite{kleman11}):
 
 In a \textit{Volterra process} the basic ingredient is a \textit{cut surface} $\Sigma$ drawn in a unstrained specimen, bordered by a closed line $\Lcal=\partial \Sigma$, whose lips $\Sigma^+$, $\Sigma^-$ are displaced one relatively to the other by a \textit{rigid} displacement ${\textbf{d}(\textbf{x})}=\textbf{b}+\bm \Omega(\textbf{x})$; $\textbf{b}$, a translation (yielding a \textit{dislocation} on $\Lcal$) and $\bm \Omega$, a rotation (yielding a \textit{disclination} on $\Lcal$) are proper elements of the symmetry group $H$ of the specimen. 
After addition of missing material in the void (if a void is created by the relative displacement) or removal of superfluous matter (if there is multiple covering), the atomic bonds are reset on the lips and the specimen let to elastically relax. 

The line defects $\Lcal$ thus obtained are \textit{quantized} if $ H$ is discrete. If the VP is performed in an amorphous material, where $\textbf{b}, \ \bm \Omega\in E(3)\sim SO(3)\ltimes\mathbb{R}^3$ are continuous elements of symmetry, the Volterra defects are \textit{continuous}.
In both cases there is no singularity left along the lips of the cut surface, after completion of the VP; $\Lcal$ is a \textit{perfect} defect line.

The VP fails if $H\notin E(3)$ is a gauge group, e.g. vortex lines in superconductors and superfluids. But there is a \textit{unique} theoretical frame in which quantized Volterra and gauge line singularities both appear as defects: 

Let $x \in M$ on which a large group $G$ is acting. The elements of $h\in G$ such that $h.x =x$ form a subgroup $H \subset G$, the \textit{little group}; the set of subgroups $g\, H \, {g} ^{-1}$ which are the little groups of $\{g.x| \ g\in G\}$ conjugated to $H$ form a conjugacy class $\left[G:H \right]$. The topological defects of $M$ 
are classified by the non-trivial elements of the homotopy group $\pi_{1}(\cal{M})$ of the coset space $\mathcal{M}=G/H \equiv G/(gHg^{-1})$.
 There are as many coset spaces attached to a group $G$ as there are conjugacy classes $\left[G:H \right]$. Each $\left[G:H \right]$ yields a possible realization of an ordered medium, whose manifold of internal states \cite{toulouse76} is precisely its coset space $\mathcal{M}$.

This \textit{topological classification of defects} extends to any physical medium with a symmetry group $G$ broken to $H$.
It also extends to defects of any dimensionality, e.g., $\pi _{2} (\cal{M})$ classifies point defects. For reviews see \cite{mermin,michel,kleman82b,trebin}. 

\textit{Cosmic forms} are the VP defects (cut hypersurface $\Sigma$, 2D \textit{world sheet} $\Lcal=\partial\Sigma$) classified by the symmetry elements $g\in P(4) \sim L_0\ltimes \mathbb{R}^{1,3}$; $L_0$ is the orthochronous connected Lorentz group, $P(4)$ the Poincar\'e group. If $g$ is an angle $\alpha \in P(4)$, it would necessarily match for some value to the quantized $\alpha$ resulting from the gauge field theory of {cosmic strings}. A cylindrically symmetric cosmic string results from a \textit{cut-and-glue process} performed in the Minkowski spacetime $M^4$, much akin to the VP for {\textit{wedge}} disclinations in CMP, i.e., disclinations  located along the rotation vector $\bm\Omega$. Other types of isometries in $P(4)$ yield other types of defects \cite{tod94,puntigam97}. In these papers the analysis is developed in the frame of the ECT, which, as demonstrated in \cite{hehl}, is the local gauge theory of the Poincar\'e group. However the defect classification is the same as that one gotten from VP's in $M^4$.  
\\ \vspace{-10pt}
 
 Sect.~\ref{poinforms} presents the classification of generic \textit{cosmic forms} including  \textit{cosmic disclinations} and \textit{cosmic dislocations}.
The subgroups of $L_0$ can be partitioned into four sets of conjugacy classes, in which one recognizes three kinds of disclinations: 

$-$ \textsf{$m$-forms}, related to continuous symmetry rotations $\alpha$,

$-$ \textsf{$t$-forms}, related to Lorentz boosts (hyperbolic rotations),

$-$ \textsf{$r$-forms}, related to null rotations, 

\noindent each kind belonging to a different conjugacy class. The fourth conjugacy class, the Lorentz group itself, contains all the forms just listed and no other.

$m$- and $t$-forms are described in the literature \cite{tod94,puntigam97}; $r$-forms have never been considered so far.
We propose the following physical interpretation. Wigner \cite{wigner39} has demonstrated the equivalence between the unitary irreducible representations of subgroups of $L_0$ and the elementary particles. Thereby, the subgroups belonging to the same conjugacy class correspond to a well-defined type of particles. 
We assign the fourth conjugacy class to an 'amorphous' substance, a \textit{vacuum}. In the cosmological literature, this vacuum is treated in the frame of the quantum theory of fields (QFT) \cite{linde05ht}; it is often identified with the cosmological constant or/and the dark matter.

In Sect.~\ref{forms} we calculate the line elements $ds^2$ of these three classes, for the \textit{wedge forms} as well as for the \textit{twist forms}. 'Wedge' has been defined above in CMP, 'twist'  refers to a line orthogonal to the rotation it carries. This terminology is perfectly acceptable for the $m$-forms, which are topologically equivalent to CMP disclinations.  The analogy goes further: a twist CMP disclination is topologically stabilized by infinitesimal dislocations or disclinations attached to it \cite{klfr08}. 
The same process occurs for twist forms in spacetime, and more generally for any non-wedge, {mixed}, segment along a curved $m$-form; it also occurs for $t$- and $r$-forms. In all those cases the singularity of the form is not along the world sheet, but along an hypersurface that surrounds the central world sheet, a \textit{world shell}.

In Sect.~\ref{disc}  
we start from the remark that  the elements of symmetry carried by $t$- and $r$-forms (in the VP sense) do not belong to the maximally symmetric group that encompasses the \textit{space} properties of isotropy and homogeneity $-$ thereby they do not obey the narrow cosmological principle (nCP), whereas the $m$-forms do. But $t$- and $r$-forms are compatible with the perfect cosmological principle (pCP) as in \cite{bondi48,hoyle48}, where the embedding spacetime is de Sitter: $r$- and $t$-forms are typical defects of a $dS_4$ vacuum.%a vacuum with a positive cosmological constant, which could be met during the process of inflation \cite{vilenkin06}. 

Finally we stress that cosmic forms can assemble into networks  which can adjust to any curvature, as disclinations do in amorphous solids % \cite{kleman79,kleman83} or in Frank and Kasper phases \cite{frank58b,frank59}, 
where they yield zero Euclidean curvature \cite{kleman79}. %Could this geometry be in direct relation to the curvature of the spacetime at the time of inflation?

 \section{Forms in a Minkowski spacetime $M^4$} \label{poinforms} 

\subsection{The classification of cosmic forms} \label{cosmicforms} 

We restrict our discussion to $M^4$; the extension to $dS_4$ is straightforward.  
\begin{figure}[h]
\begin{center}
\includegraphics[width=2. in]{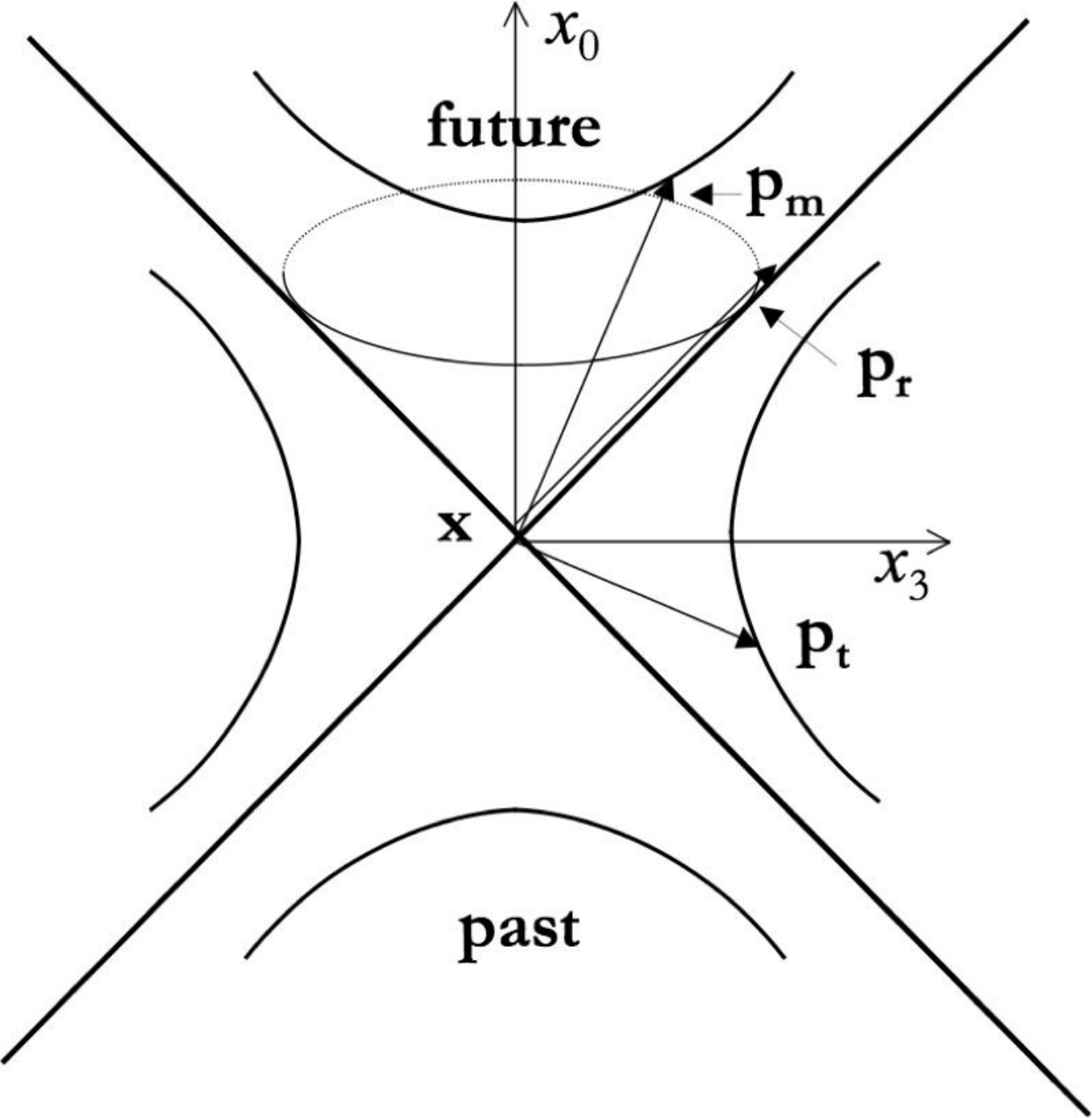}
   \caption{ 3D representation of the light cone at some event $\textbf{x}$, cut in the $\mathbf{e}_0\,\mathbf{e}_3$ plane. 
   The three non-trivial orbits of the extremities of the vectors ${p}^a_m$ (timelike), ${p}^a_r$ (null), ${p}^a_t$ (spacelike) are respectively hyperboloids of two sheets, the light cone with the origin excluded, hyperboloids of one sheet. A cross section of the future light cone is sketched in perspective; it is a 2-sphere in 4D. The origin ${p}^a=0$ is a 4$^{rth}$ orbit in itself.}
 \label{fig01} 
 \end{center}
\end{figure}

$P(4)$ is partitioned into four conjugacy classes, yielding four corresponding coset spaces. Michel \cite{michel} obtains these classes by considering the action of  $P(4)$ on 4-momenta $p^a$, each orbit corresponding to a different norm of $\|{p^a}\|^2$, each stratum to a set of orbits corresponding respectively to timelike, spacelike, null, and vanishing momenta, Fig.~\ref{fig01}. %The use of the group action on 4-momenta can be given a cosmological interpretation, at least for timelike and null momenta, within the scope of Wigner's theory \cite{wigner39}. %, see Sect.~\ref{stab-cryst}.
\\ \vspace{-10pt}

 \textit{1- Timelike 4-momenta} 
 
    Let ${p}_m^a = \{{p}_m^0,{p}_m^1,{p}_m^2,{p}_m^3\}$ be some timelike 4-momentum. The little group that leaves ${p}_m^a$ invariant is isomorphic to the group of 3D rotations; $H_m\sim {SO(3)}$. 
    The orbit of ${p}_m^a$ belongs to a hyperboloid of two sheets. In Fig.~\ref{fig01}, where $p^{0}>0$, it is the sheet to the future of $\textbf{x}$. 
    The coset space $\mathcal{M}_m =L_0/SO(3)$ is topologically equivalent to the orbit, i.e., $\mathcal{M}_m=H^3$. 
The corresponding homotopy groups are all trivial
\begin{equation}
\label{1}
  \pi_{i}({H^3)\sim \{1\}}.
\end{equation} 

We are left with continuous Volterra defects, \textit{$m$-dislocations} and \textit{$m$-disclinations} carrying elements of the full little group of symmetry, translations and rotations included, $ P_{m} \sim H_m\ltimes \mathbb{R}^{1,3} \sim SO(3)\ltimes\mathbb{R}^{1,3}$. \\ \vspace{-10pt}
 
 \textit{2- Spacelike momenta} 

The little group that leaves a spacelike 4-momentum ${p}^a_t$ invariant is $H_t \sim {PSl(2,\mathbb{R}) \sim Sl(2,\mathbb{R})/\mathbb{Z}
_2}$  \cite{sternberg94}. The coset space $\mathcal{M}_t$ is topologically equivalent to a spacelike hyperboloid of one sheet $\mathcal{M}_t\sim S^2 \times \mathbb{R}$; thus:
 \begin{equation}
\label{2}
  \pi_{i}(\mathcal{M}_{t})\sim \pi_{i}(S_{2}) \rightarrow \\ 
   \pi_{1}(\mathcal{M}_{t})\sim \{1\},\,\pi_{2}(\mathcal{M}_{t})\sim \mathbb{Z},\,\pi_{3}(\mathcal{M}_{t})\sim \mathbb{Z},\,\pi_{4}(\mathcal{M}_{t})\sim \mathbb{Z}_2.
\end{equation}
There are Volterra continuous and topologically stable defects, whose dimensionalities in $M^4$ are: 1 for $\pi_{2}(\mathcal{M}_{r})$ and 0 for $\pi_{3}(\mathcal{M}_{r})$; $\pi_{4}(\mathcal{M}_{r}) $ designates possible \textit{configurations} (in the sense of \cite{kleman78}).
All line defects are continuous. By definition, they obtain by a VP carrying an element of the full little group $P_{t}\sim H_t\ltimes \mathbb{R}^{1,3}$. \\ \vspace{-10pt}

     \textit{3- Null momenta} 
     
     The little group that leaves ${p}_r^a$ invariant is isomorphic to the 2D Euclidean group; $H_r\sim {E}(2)= SO(2)\ltimes\mathbb{R}^2 $, the semidirect product of a circle by a plane.
The (unique) orbit of ${p}_r^a$, which is also the coset space $\mathcal{M}_r = L_0/E(2)$, is  topologically equivalent to the future null cone with apex removed, $ \|{p}_r^a\|^2=0$ $\mathcal{M}_r\sim S^2 \times \mathbb{R}$.  One gets the same homotopy as for $\mathcal{M}_t$, namely: 
 \begin%{multline}
{equation}
\label{3}
  \pi_{i}(\mathcal{M}_{r})\sim \pi_{i}(S_{2}) \rightarrow \\ 
   \pi_{1}(\mathcal{M}_{r})\sim \{1\},\,\pi_{2}(\mathcal{M}_{r})\sim \mathbb{Z},\,\pi_{3}(\mathcal{M}_{r})\sim \mathbb{Z},\,\pi_{4}(\mathcal{M}_{r})\sim \mathbb{Z}_2.
\end%{multline}
{equation}
Hence there are both Volterra continuous and topologically stable defects, whose dimensionalities in $M^4$ are: 1 for $\pi_{2}(\mathcal{M}_{r})$ and 0 for $\pi_{3}(\mathcal{M}_{r})$; $\pi_{4}(\mathcal{M}_{r}) $ designates possible \textit{configurations}.
All line defects are continuous defects. By definition, they obtain by a VP carrying an element of the full little group $P_{r}\sim H_r\ltimes \mathbb{R}^{1,3}$.\\ \vspace{-10pt}
  
\textit{4- Vanishing momenta} 

$p^a=\{0,0,0,0\}$: the little group is the full proper Lorentz group $L_0$; the coset space $\mathcal{M}_{v} = L_0/L_0$ is reduced to a point. Therefore all the homotopy groups $\pi_i(\mathcal{M}_{v})$ are trivial and the only possible defects are continuous VP defects. All the elements of $L_0$ have been scanned with the previous little groups, and thus this case embraces all the forms already cited. 

 \subsection{The significance of topological defects: $M^4$-crystals, vacuum} \label{stab-cryst} 
  
\indent \indent  \textit{1$-$ The first three conjugacy classes}.  In the spirit of \cite{wigner39}, the relationship between elementary particles and conjugacy classes of $L_0$ subgroups suggests to identify a set of identical elementary particles $-$ of the same 4-momenta ${p}^a$ in $M^4$, of constant density, represented by some subgroup $H$ of $L_0$ leaving ${p}^a$ invariant, forming a \textit{congruence of geodesics} along ${p}^a$, $-$ to a \textit{crystal in $M^4$}, whose symmetry group is precisely the little group that leaves $p^a$ invariant. 
 Thus the first three conjugacy classes yield three types of crystals, with inhomogeneous symmetry groups $P_m$, $P_t$, and $P_r$ \footnote{We leave aside the question of the physical reality of tachyons.}. Thereby the topological and Volterra defects of these crystals would be those defined in the previous section. The use of 4-momenta to getting the three types of continuous forms is therefore fully justified.  
  
\textit{2$-$ The fourth conjugacy class} appears here in a new perspective, not so trivial. 
The $m$-, $r$- and $t$-forms it embraces could be observed if there is some substance that possesses all the $L_0$ symmetries, a sort of spacetime amorphous substance. We tentatively identify this substance with a QFT vacuum of a de Sitter spacetime; the de Sitter spacetime is endowed with a positive cosmological constant.

 \subsection{The cosmological principle} \label{cosmprinc}  
 
It is worth analyzing this 'crystal' classification with regard to the cosmological principle, which
states that any two elements belonging to the same spacelike slice are equivalent, at least at some large enough scale.
This \textit{narrow} cosmological principle (nCP) is in contrast with
a \textit{perfect} cosmological principle (pCP) which states that any two events in spacetime are equivalent \cite{bondi48,hoyle48}.
Expressed in terms of symmetry groups, a)- nCP states that the spacetime is foliated by maximally symmetric spacelike hypersurfaces parameterized by the cosmic time, with Volterra defects related to their isometry groups ($nCP \, \sim SO(3)\ltimes\mathbb{R}^{3}$ in $M^4$); b)- pCP implies a physical situation invariant under the Poincar\'e group $pCP \sim P(4)$ (in $M^4$) or the de Sitter group $pCP \sim SO(1,4)$ (in $dS_4$). 
  
  The only nCP-compatible forms are the $m$-forms. This does not preclude the existence of $r$- and $t$-forms, under the status of imperfect defects. On the other hand all three types of forms are pCP-compatible, which emphasizes the QFT vacuum hypothesis. pCP defects are not valid physical defects at the present epoch, but may have been associated with a de Sitter universe in the primeval Universe.
    
  \section{Continuous cosmic forms} \label{forms}

We center our discussion on \textit{disclinations} and their $ds^2$ in $M^4$. Energetic and dynamical properties, the detailed structure of the core, are left aside. %\textit{Dislocation} structural properties vary much from one type of spacetime to another. Contrarywise
Disclinations being related to the $L_0$ subgroup, present in all spacetimes, the following results are valid \textit{mutatis mutandis} for any spacetime.
\textit{Dislocations} play a distinctive role in $M^4$, in all processes that involve movement, change of shape, relaxation, etc. of disclinations, somewhat analogously to their role in CMP \cite{klfr08}. 

The main concepts related to cosmic forms will be developed for the $m$-form %(the classical cosmic string of the present literature), 
 and will not be detailed again for the $t$- and the $r$-forms. 
 
We use the spin representation of the Lorentz group 
$S\ell(2,\mathbb{C})\rightarrow {L_0}$. $S\ell(2,\mathbb{C})= \widetilde{{L_0}}$, the covering group of ${L_0}$, is the group of 2$\times$2 matrices 
$A=\begin{array}{|cc|}
  a    &   b \\
   c   &   d \\
\end{array}$
with complex entries and unit determinant $ad-bc=1$.  Any element $g \in  {L_0}$ lifts to two elements $A, -A \in S\ell(2,\mathbb{C})$. A short reminder of the $S\ell(2,\Cc)$ representation can be found in \cite{kleman11}. 

 \subsection{General comments about cosmic forms}\label{gforms}
 
 We summarize the main results that emerge in this section:

 $-$ there are \textit{wedge} and \textit{twist} cosmic forms.  A \textit{mixed} $m$-form is a generic non-wedge form,
 
 $-$ the 2-plane world sheet $\Lcal$ of a \textit{wedge} cosmic $m$-, $t$- or $r$-form is invariant (locally or globally) under the action of $g \in  {L_0}$; it carries the singularity of the wedge form,
  
$-$ the world sheet $\Lcal$ of a \textit{twist} cosmic $m$-form is  a 2-plane orthogonal to the 2D rotation axis; $\Lcal$ is not invariant under the action of $g$. The singularity is carried by a 3D timelike manifold, the \textit{world shell} $\Scal$ whose size varies along $\Lcal$,
     
$-$ the world shell of a twist or mixed $m$-form can be stabilized at a constant size by the addition of a density of infinitesimal cosmic forms (dislocations or disclinations) attached all along the $m$-form; a $m$-form $\Lcal$ can curve to any shape under the effect of a variable attachment of dislocations or disclinations all along thereby. There is a comparable property in CMP \cite{klfr08},

 $-$ an analogous analysis applies to $t$- and $r$-forms, but with differences stemming in the more subtle nature of the rotation angles (hyperbolic and null angles, respectively),
     
$-$ we are concerned with forms characteristic of some substance inhabiting the spacetime, for example a set of particles defined by a constant 4-momentum $p^a$ with little group $H_a \subset L_0$. If $A $, the symmetry element carried by the cosmic form, belongs to $ \widetilde{H_a}$, then the cut hypersurface $\Sigma$ is not observable; it is a \textit{perfect} form. A cosmic form is \textit{imperfect} when the symmetry element  $A_u$ it carries $ \notin\widetilde{ H_a}.$; the cut hypersurface $\Sigma_u$ is made visible.

 \subsection{$m$-cosmic forms}\label{mforms}
    
     \textit{1- The wedge $m$-form}
     
      In the sequel, $\Lcal$ is taken along
  the $\{ \textbf e_0 \,\textbf e_3 \}$ plane. We do not consider the (mathematically feasible) situation where the singularity is spacelike, as in \cite{puntigam97}, in which case the form is a hypersurface in a 3D space, not a line. \\ \vspace{-10pt}

 Let $p_m^a $ be a timelike 4-momentum, invariant about any rotation $g \in H_m \sim SO(3)$ that leaves invariant the spacelike manifold ${P}_{m\bot}$ orthogonal to $p_m^a $. Hence $g$ is a global rotation of ${P}_{m\bot}$ about some rotation axis $\bm \varpi  \subset {P}_{m\bot}$; the 2-plane $\Omega \equiv \{ \textbf p_m^a  \,\bm \varpi \}$ is a 2-plane of rotation.
 
To make things simple, take $p_m^a = \{p_m^0,0,0,0\}$; hence ${P}_{m\bot} \equiv \{ \textbf e_1 \,\textbf e_2 \,\textbf e_3 \} $. Any element of $\widetilde{H_m}$ has the form $A_m=\scriptsize{\begin{array}{|cc|} a & b\\  -\bar{b} & \bar{a} \end{array}} \in SU(2) \sim \widetilde{SO(3)}$ \cite{wigner39}. The displacement of a point
$X_{m}^a=\{X_m^0,X_m^1,X_m^2,X_m^3 \}$ on the cut hypersurface $\Sigma_m$ 
reads $d_m =x_{m}-X_m = A_m \, X_{m} \,A_m^*-X_m$. In this equation, $X_{m}^a$ is represented by the 2$\times$2 hermitian matrix  ${{X}}
=\begin{array}{|cc|}
 X^0+X^3     &  X^1-i\,X^2  \\
  X^1+i\,X^2    &   X^0-X^3
\end{array}.$
We also assume $a=\exp- i\alpha/2,\,b=0$ (which yields $\{\bm \varpi \} =\{ \textbf e_3\}$);
  $\Lcal \equiv \{\textbf e_0\,\textbf e_3\}$, the plane of rotation is \textit{locally invariant} under the action of $A_m$: 
  \begin{equation}
\label{4}
{x}_m^0=X_m^0,\quad {x}_m^3 = X_m^3,\quad {x}_m^1+i\,{x}_m^2 =({X}_m^1+i\,X_m^2)\exp{i\alpha}. 
\end{equation} 
$ \{\textbf e_3 \}$ is the axis of rotation in the spacelike submanifold $P_{m\bot} $, along which direction lies the singularity of the wedge form in $P_{m\bot} $; $\{\textbf e_0 \}$ is the axis of rotation in the timelike submanifold $\{ \textbf e_0\,\textbf e_1\,\textbf e_2 \}$. 
  
  This {wedge} $m$-form is the same as the conical cosmic string of ref. \cite{vilenkin81b,hiscock85,puntigam97,tod94}. All the displacements take place in a 3-space orthogonal to $p_m^a $; the analysis of the form in spacetime does not differ from the analysis of a continuous disclination in the habit Euclidean 3-space of an amorphous substance. Since wedge disclinations in space can curve to any shape, similarly $m$-disclination loops can take any shape; however a full understanding of this property requires the notion of twist disclination, below.\\ \vspace{-10pt}

The line element of a straight $m$-disclination obtains by applying the Volterra process to the spacetime \cite{tod94}. In the VP the coordinates of a undeformed Minskowski spacetime $\{T,R, \phi,Z\}$ are transformed to: 
$\{t=T, \quad r = R, \quad \phi =  \frac{\Phi}{1 - \frac{\alpha}{2\, \pi} },\quad z = Z,\}$
 when running from $\Phi = 0$ to $\Phi = 2\, \pi - \alpha$.  Substituting into the line element for an undeformed $M^4$ one gets:
\begin{equation}
\label{5}
ds^2 = -dt^2 +dz^2 + dr^2 + a^2\,r^2 \,d\phi^2, \qquad a =1- \frac{\alpha}{2\ \pi}.
\end{equation}
 The Riemann tensor vanishes everywhere in the spacetime, except on $\Lcal$.
 
Eq.~\ref{5} can also be written:
\begin{equation}
\label{6}
ds^2 = -dt^2+(dx+\frac{\alpha}{2\,\pi}\,y \,d\phi)^2 + (dy-\frac{\alpha}{2\,\pi}\,x \,d\phi)^2 +dz^2,
\end{equation} where the rotation $\alpha$ about the $\{\textbf e_3\}$ axis is apparent.\\ \vspace{-10pt}

There are two types of disclinations:

$-$ {$\alpha >0, \, a < 1$}; a sector of angle $2\, \pi \,(1-a)$ is cut out from the spacetime and the resulting edges are identified; $\alpha$ is a \textit{deficit angle}. In the CMP terminology, this is a \textit{positive disclination}. All cosmic strings are of this type,

$-$ {$\alpha < 0, \, a > 1$}; a sector of angle $2\, \pi \,(a-1)$ is inserted and its edges are identified with the lips of the cut. In the CMP terminology, this is a \textit{negative disclination}.\\ \vspace{-10pt}

The mass-energy $\mu$ per unit cosmic line length can be written
$\mu = \frac{c^2}{4\,G}\,({1-a})\equiv   \frac{c^2}{8\,\pi \,G} \,\alpha,$
$G$ is the gravitation constant, $c$ 
the light velocity \cite{langer70}. 
The positivity of $\mu$ forbids the existence of negative disclinations. However this is without taking into account the possibility of multiple vacua of positive as well as negative energies in inflation theory \cite{vilenkin06}.
\\ \vspace{-10pt}

 \textit{2- The twist $m$-form; its world shell}

Consider a $m$-form whose 2-axis of rotation is $\{\textbf e_0 \, \textbf e_2\}$, with the same world sheet $\Lcal \equiv\{\textbf e_0 \, \textbf e_3\}$ as above. $\Lcal$ is not invariant, even globally, under the action of the VP. In the 3 space $\{\textbf e_1 \, \textbf e_2 \,\textbf e_3\}$ the cosmic form is along the $\{\textbf e_3\}$ axis, and the rotation about the $\{\textbf e_2\}$ axis, orthogonal to it. We keep the CMP terminology of \textit{twist disclination}.

When navigating around $\Lcal$, the old coordinates coordinates $\{X^1,X^3\}$ become:
\begin{equation}
\label{7}
 x = X^1 \,\cos \frac{\alpha}{2\, \pi}  \, \phi - X^3 \,\sin\frac{\alpha}{2\, \pi}  \, \phi  , \quad z= X^1 \,\sin \frac{\alpha}{2\, \pi}  \, \phi + X^3 \, \cos \frac{\alpha}{2\, \pi}  \, \phi ,  \quad r=R \quad \phi = \Phi, \end{equation} 
and the line element takes the form:
\begin{equation}
\label{8}
ds^2 = -dt^2 +(dx+\frac{\alpha}{2\, \pi} \,z \,d\phi)^2 + dy^2 +(dz - \frac{\alpha}{2\, \pi} \,x \,d\phi)^2.
\end{equation}
The singularity is not carried by $\Lcal$ but by a timelike \textit{world shell} $\Scal_m \equiv \{\Hcal_m=0\}$: 
\begin{equation}
\label{9}
\Hcal_m=r-\frac{\alpha}{2\, \pi}\,z\,\sin\phi.
\end{equation} 
The non-vanishing Christoffel symbols, inverse metric components, all behave as  $\Hcal_m^{-1}$ and thus diverge on $\Scal_m.$   There are incomplete geodesics that terminate (or start) on $\Scal_m$. The metric of Eq.~\ref{8} being valid in the inner part of $\Scal_m$ as well as the outer part, there are such geodesics on both sides of $\Scal_m$. An  equivalent property has not been pointed out in CMP.

The appearance of a singular world shell has some mathematical advantages. It is argued in \cite{geroch87} that postulating a Dirac distribution of curvature on a 2D manifold is not a well-posed mathematical problem, but that there is no such trouble for a 3D one. \\ \vspace{-10pt}

 \textit{3- Relation between twist cosmic $m$-forms and dislocations}
   
\begin{figure}[h]
\begin{center}
 \includegraphics[width=2.8 in]{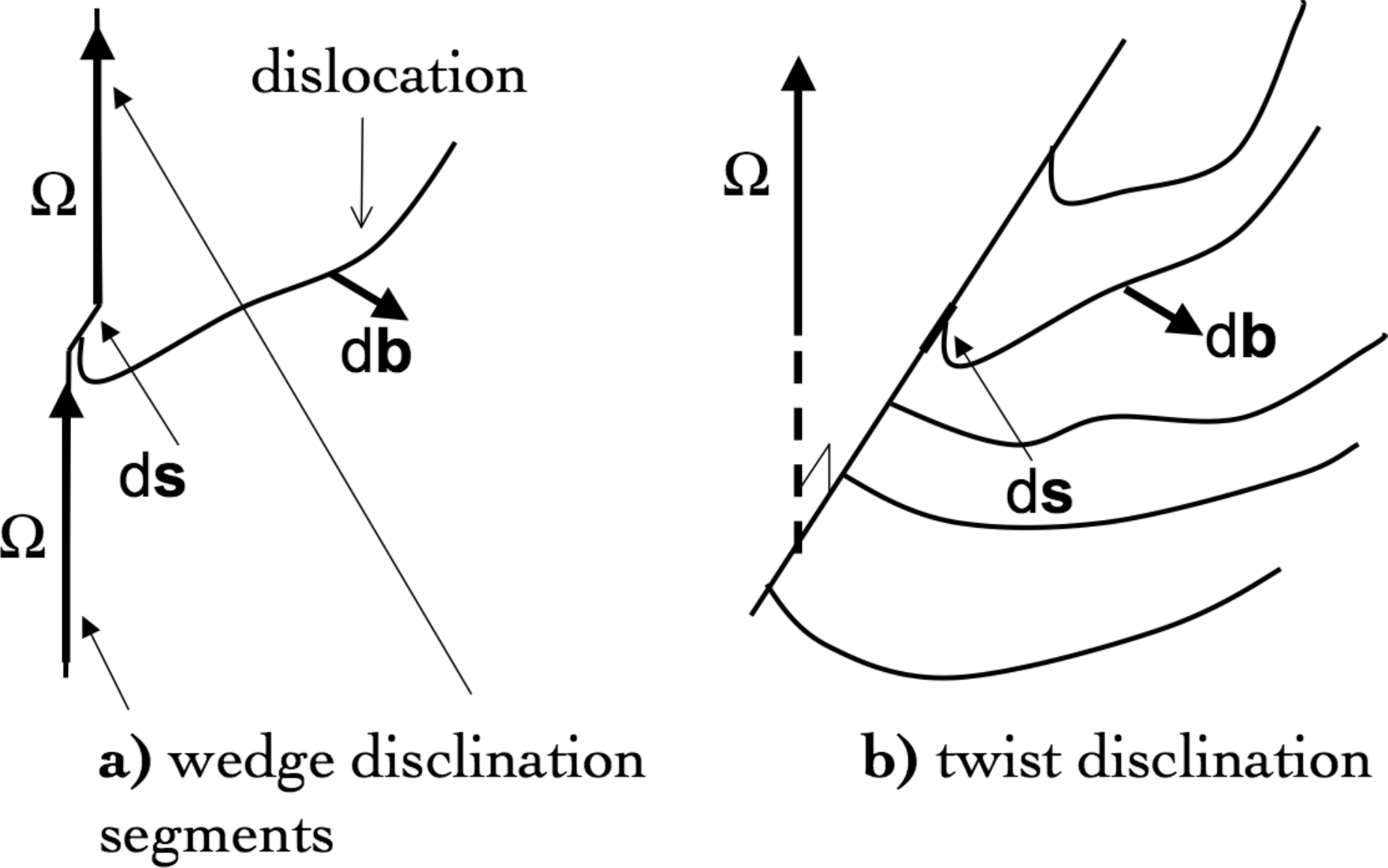}
   \caption{a)- Kink $d\textbf s$ along a CMP wedge disclination of strength $\bm \Omega$; b)- twist disclination made of an assembly of such kinks with a constant Burgers vector density $d\textbf b / d\textbf s$.}
 \label{fig2} 

\end{center}\end{figure}

  Twist disclinations and dislocations are topologically related objects in CMP. This is illustrated Fig.~ \ref{fig2}a, where a wedge disclination line is kinked along a segment $d\textbf s$ orthogonal to the VP rotation $\bm \Omega$; $d\textbf s$ is a twist segment along a wedge line. Clearly, any point M on the cut surface of such a defect does not suffer the same displacement $\bm \Omega \times \textbf{AM}$ (we assume $|\bm\Omega|$ small), whether A is above the kink $\textbf{d}^+=\bm \Omega \times \textbf{A}^+\textbf M$ or below $\textbf{d}^-=\bm \Omega \times \textbf{A}^-\textbf M$ ($\textbf{d}^+$ and $\textbf{d}^-$ do not depend on the position of $\textbf{A}^+$ and $\textbf{A}^-$), and the difference $\textbf{d}^+ -\textbf{d}^- = \bm \Omega \times d\textbf s$ is a constant. This results in the presence of a dislocation of Burgers vector:
    \begin{equation}
\label{10}
d\textbf b = \bm \Omega \times d\textbf s.
\end{equation}
attached to the kink, see \cite{klfr08}. A twist disclination is nothing else than a set of such kinks, i.e., attached dislocations with a line density $\bm \Omega \times \frac{d\textbf s}{ds}$, Fig.~ \ref{fig2}b.

  \indent Analogous topological relations exist between cosmic forms and dislocations in spacetime. Consider an \textit{edge} dislocation, world sheet $\{\textbf e_0 \, \textbf e_3\}$, Burgers vector $\textbf b$ along the $\{ \textbf e_1\}$ direction. The line element is:
\begin{equation}
\label{11}
ds^2 = -dt^2+(dx+\frac{b}{2\,\pi}\,d\phi)^2 +dy^2 +dz^2. 
\end{equation}  Again, it appears a world shell $\Scal_{e} \equiv \{\Hcal_{e} = r-\frac{b }{2\,\pi} \, \sin \phi = 0\}$) along which 
some Christoffel symbols, inverse metric components, and  geodesic equations behave as  $\Hcal_{e}^{-1}$ and thus diverge (cf. an analogous result in \cite{krasnikov07}, in a different context). $\Scal_{e}$ size is approximately $|b|/(4 \, \pi)$. 

Compare $\Hcal_m=r-\frac{\alpha }{2\,\pi}\,z \,\sin \phi$ and $\Hcal_{e_x}=r-\frac{b }{2\,\pi} \, \sin \phi$; these two expressions can be made equal along successive segments $\delta z= b/\alpha$: hence it is possible to attach a set of dislocations of Burgers vector $b =\alpha \ \delta z$, one per length $\delta z$, and thereby get rid of the variation of $\Scal_{m}$ along the twist form. In other words (passing to a continuous Burgers vector) the twist form is dressed with a dislocation density $db/dz = \alpha$.
 Eventually the core configuration of the twist $m$-form is translationally invariant along the $\{ \textbf e_3\}$ axis. 

We do not investigate the detailed configuration of the attachment and the energy of the core, a topic in itself.

\subsection{$t$-cosmic forms} \label{hypdiscli} 

 \indent \indent\textit{1- The wedge $t$-form} 

Consider a VP carrying the Lorentz boost $A_t=\begin{array}{|cc|}
 e^{\frac{\omega}{2}}     &  0  \\
 0    &    e^{-\frac{\omega}{2}} 
\end{array} =A_t^*, \, A_t \in \widetilde{H_t}$.
 The spacelike directions invariant under the action of $A_t$ can be written $p^a_t = \{0,p^1,p^2,0\}$, $\forall \, p^1,\, \forall \,p^2$;
 a pair of spacelike directions $\{0,p^1,0,0\},\ \{0,0,p^2,0\}$ defines a \textit{locally invariant} 2-plane which could be taken as a world sheet for a wedge $t$-form carrying $A_t$. But this plane is spacelike and does not yield a line defect in space. On the other hand a timelike 2-plane like 
$\Lcal = \{\textbf e_0 \, \textbf e_3\}$, which is \textit{globally invariant}, provides a suitable 'wedge' world sheet.  \\ \vspace{-10pt}

The same method that yielded Eq.~\ref{7} extends to getting the line element of this $t$-form. First we have:
\begin{equation}
\label{12}
 t = T \,\cosh \varpi  \, \Phi + Z \,\sinh \varpi  \, \Phi  , \quad z= T \,\sinh \varpi  \, \Phi+ Z \, \cosh \varpi  \, \Phi ,  \quad r=R \quad \phi = \Phi \end{equation} 
 where $ \varpi = \frac{\omega}{2 \, \pi}$,
($\varpi  \, \phi$ varies between $0$ and $\omega$ when traversing a loop that surrounds $\Lcal$). Then:
\begin{equation}
\label{13}
ds^2 = -(dt - \varpi \,z \,d\phi)^2+dr^2 + r^2 \,d\phi^2 +(dz - \varpi \,t \,d\phi)^2,
\end{equation}

\noindent as in  \cite{tod94,puntigam97}. The Riemann tensor attached to this line element vanishes identically. Tod \cite{tod94} has shown that this disclination carries singularities (distributional curvature and torsion) in the $\Lcal$ plane. Thus we still call it a wedge form.\\ \vspace{-10pt}

 \textit{2- The twist $t$-form; attached dislocations}\label{coret}

Consider a $t$-form (''disclination 7'' in \cite{puntigam97}) with the same $\Lcal \equiv \{\textbf e_0 \, \textbf e_3\}$ as above, whose hyperbolic rotation is in the $\{\textbf e_0 \, \textbf e_2\}$ plane. $\Lcal$ is not invariant in the VP, even globally. In the 3-space $\{\textbf e_1 \, \textbf e_2 \,\textbf e_3\}$ the cosmic form is along the $\{\textbf e_3\}$ axis, and the hyperbolic rotation along the $\{\textbf e_2\}$ axis, orthogonal to $\{\textbf e_3\}$. We keep the terminology of \textit{twist} disclination.

The variation of the boost when navigating around $\Lcal$ is:
\begin{equation}
\label{14}
 t = T \,\cosh \varpi  \, \Phi + Y \,\sinh \varpi  \, \Phi  , \quad y= T \,\sinh \varpi  \, \Phi+ Y \, \cosh \varpi  \, \Phi ,  \quad r=R \quad \phi = \Phi, \end{equation} 
and the corresponding line element is (notice a correction of sign with respect to \cite{puntigam97}):
\begin{equation}
\label{15}
ds^2 = -(dt - \varpi \,y \,d\phi)^2+(dy- \varpi \,t \,d\phi)^2 + dx^2 +dz ^2.
\end{equation} 

The Riemann tensor vanishes identically. In analogy with the $m$-twist form, the inverse metric and the Christoffel symbols diverge on a hypersurface $\Scal_t \equiv \{\Hcal_t =0\}$:\begin{equation}
\label{16}
\Hcal_t = r - \varpi \,t\,\cos \phi.
\end{equation} 

$\Scal_t$ is not time-invariant. In analogy with the twist $m$-case, this can be cured by a suitable dislocation density attached to the $t$-form. The relevant dislocations have the line element:
\begin{equation}
\label{17}
ds^2 = -dt^2+dx^2 +(dy+\frac{b_y}{2\,\pi}\,d\phi)^2 +dz^2. 
\end{equation}  The related singular hypersurface is $\Scal_{e_y} \equiv \{\Hcal_{e_y} = r-\frac{b_y }{2\,\pi} \, \cos \phi=0\}$.
The attached dislocation density is $db_y/ dt = \varpi$. 

\subsection{$r$-cosmic forms}\label{forms-light}

The general expression for an element $A_r$ of $\widetilde{H_r}$ which leaves locally invariant the null direction $k^a =\{1,0,0,1\} \propto p_r^a =\{p,0,0,p\}$ can be written: 
$A_r= \begin{array} {|cc|}
  \exp{-i\frac{\alpha}{2}}   \, \,\,&\,  -\bar{\sigma}\,\exp{i\frac{\alpha}{2}} 
  \\ 0 \,\,&\,   \exp{i\frac{\alpha}{2}} \end{array},\ \mathrm{with} \ \sigma = \sigma_1 + i\,\sigma_2,$  \cite{wigner39}.
No other null direction is invariant under the action of $A_r$. Therefore there is no world sheet \textit{locally} invariant in any VP of the $A_r$ type. On the other hand one can show that there is one 2-plane (and only one) that is VP \textit{globally} invariant. Choosing this plane as the world sheet would define a wedge $r$-form.
  The null 3-manifold $P_{r\bot} \equiv x^0-x^3 = 0$ orthogonal to $k^a$ is globally invariant under the same action.   We haven't carried out the rather cumbersome calculation of the line element of this wedge $r$-form. This is in contrast with the $m$-case, where a 3D spacelike manifold is invariant; in that sense, a $r$-disclination does not obey the cosmological principle. The (rather complicated, but never considered up to now) establishment of the line elements of two twist types of $r$-forms, as well as a discussion of the attached dislocations on one example, can be found in the Appendix. \\ \vspace{-10pt}

\subsection{Observability of the cut hypersurface of a cosmic form}\label{obscut}

This section concerns the $M^4$-'crystals' of Sect.~\ref{stab-cryst}. Consider e.g., a $m$-form, VP-constructed on a cut hypersurface $\Sigma_m$, breaking the symmetry element $A_m=\scriptsize{\begin{array}{|cc|} a & b\\  -\bar{b} & \bar{a} \end{array}}.$ $A_m$ leaves invariant any timelike 4-momentum as 
\begin{center}
$p_m^a = \{p_m^0,0,0,p_m^3\}, \quad \forall p_m^0, \, \forall p_m^3, \quad ||p_m^a||^2<0,$
\end{center}
the resulting $m$-form is a \textit{perfect} VP defect in a spacetime inhabited by the $p_m^a$ 'crystal'. For any other $p$ crystal with $p\neq \lambda p_m^a,\ \lambda = cst$, the $m$-form differs from a perfect form by the appearance of a \textit{misorientation} from one side of $\Sigma_m$ to the other. Compare indeed $p$ on one side to $p' = A_m \,p\, A_m^*$ on the other for, e.g., $p=p_r^a=\{1,1,0,0\}$; one gets $ p'_r=A_m \,p_r\, A_m^* =\{1,\cos \alpha,\sin \alpha,0\}$. $\Sigma_m$ is made visible by this misorientation; its observability gives evidence of an \textit{imperfect} form. The cut hypersurface gets a physical status, similar to a grain boundary in CMP. The
$p_m^a$ field is continuous through the cut surface, but not on the world sheet or the world shell. Since $p_m^a$ does not suffer any singularity on $\Sigma_m$, the cosmic form is not sensitive to the exact location of the cut surface. 
These arguments extend to any cosmic form carrying a symmetry element $A_t$ or $A_r$.

\section{Discussion} \label{disc}

 \indent \indent\textit{1- additional comments}
 
     \indent $-$      
     the relation between VP defects in $dS_4$ and $M^4$ can be made more precise: a \textit{contraction} \cite{inonu53} of  the group of isometries $SO(1,4)$ of $dS_4$ to $P(4)$, with respect to the little group $SO(1,3) \sim L_0$ at a point $x \in dS_4$, transforms $L_0$ into $\mathbb{R}^{1,3}$.
      Thereby the rotations in $SO(1,3)$ turn into the translations in $\mathbb{R}^{1,3}$, the corresponding disclinations into dislocations; the attachment of a dislocation in $M^4$ results from the attachment of a disclination in $dS^4$. 
     
      \indent $-$ a variable attachment density results in a curved mixed form. The attached infinitesimal densities, which are free to spread out through spacetime, constitute a topological contribution to the spacetime continuous distortions. 
      
      \indent $-$ The attachment of infinitesimal defects does not depend whether the main disclination is of a continuous or discrete strength; it applies to  cosmic strings. The dynamics of a string or a form requiring some shape fluctuation of the line defect, thereby it is attended by variable attached defect densities. 
           
      \indent $-$ in all the examples above, the Riemann curvature of an isolated wedge or twist form (before stabilization by attached densities) is vanishing (in $M^4$). We conjecture that this is always the case. This is no longer true in the presence of attached densities.\\ \vspace{-10pt}

   \indent \indent\textit{2- a few conjectural remarks}
   
\indent $-$ As already mentioned, the $m$-, $t$- and $r$-forms obey the perfect cosmological principle (pCP); $t$- and $r$-forms are not compatible with the narrow cosmological principle (nCP).
The steady state spacetime of ref. \cite{bondi48,hoyle48}, namely $dS_4$, is built precisely in order to obey pCP: its positive cosmological constant is usually interpreted as a \textit{false vacuum}; if so $r$- and $t$-forms are forms typical of a false vacuum. They could be met, as well as $m$-forms, during the process of inflation \cite{vilenkin06}, as long as the spacetime is $dS_4$; as the decay of the false vacuum proceeds in function of time $t$, the spacetime takes a generic FRW structure and the $r$- and $t$-forms are no longer VP compatible with this structure: they might acquire a status of imperfect forms, or smoothly disappear, due to their continuous character. 

 \indent \indent $-$ 
 Wedge $m$-forms are reminiscent of Regge's \textit{bones} \cite{regge61}, which carry continuous deficit angles. Bones form a \textit{skeleton} that approximate spacetime curvature, with Bianchi relations at the \textit{joints}. Similarly, one recognizes in CMP the existence of disclination \textit{networks} (skeletons), with Kirchhoff relations at the \textit{nodes} (joints), but also of \textit{conjugated disclination networks}, which are believed to be present in amorphous systems \cite{kleman83}: each network is made of disclination segments which carry deficit angles of the same sign (they cannot annihilate mutually), with opposite signs for the two networks. As a result the curvatures cancel on the average; the amorphous material can thereby live in a flat Euclidean space (as it must), at the expense of internal stresses.
   
 We speculate that cosmic forms gather into networks, with a richer structure than Regge's skeletons since they may incorporate $r$- and $t$-forms. Such networks would play a role in the setting up of the space curvature. This curvature almost vanishes at the end of the period of inflation, without the intervention of disclinations according to the present views; this might also happens before inflation or at its beginning, by the nucleation of conjugated networks that could achieve a decrease of the occupied volume hence a decrease of the local energy (at constant false vacuum energy density). Thus, the inflation process would act on a already flat space.\\

\textbf{\Large Appendix: $r$-cosmic forms} \label{forms-light}\\ \vspace{-10pt}\normalsize

%The general expression for an element $A_r$ of $\widetilde{H_r}$, which leaves locally invariant the null direction $k^a =\{1,0,0,1\} \propto p_r^a =\{p,0,0,p\}$, can be written \cite{wigner39}: 
%$A_r= \begin{array} {|cc|}
%  \exp{-i\frac{\alpha}{2}}   \, \,\,&\,  -\bar{\sigma}\,\exp{i\frac{\alpha}{2}} 
 % \\ 0 \,\,&\,   \exp{i\frac{\alpha}{2}} \end{array},\quad \mathrm{with} \quad\sigma = \sigma_1 + i\,\sigma_2.$
%No other null direction is invariant under the action of $A_r$. Therefore there is no world sheet \textit{locally} invariant in any VP characterized by  $A_r$. On the other hand one can show that there is one 2-plane (and only one) that is VP \textit{globally} invariant; this plane contains of course the null direction $k^a$. Choosing this plane as the world sheet would define a wedge $r$-form, where in analogy with a wedge $t$-form the world sheet is only globally invariant.
 % The null 3-manifold $P_{r\bot} \equiv x^0-x^3 = 0$ orthogonal to $k^a$ is globally invariant under the same action. This is in contrast with the $m$-case, where a 3D spacelike manifold is invariant; in that sense, a $r$-disclination does not obey the cosmological principle. We haven't carried out the calculation of the wedge $r$-form metric, which is rather cumbersome.\\ \vspace{-10pt}

Let $X^a =\{X^0,X^1,X^2,X^3\}$ be an event on the cut hypersurface (not defined yet) in $M^4$;
 its transform under $A_r= \begin{array} {|cc|}
  \exp{-i\frac{\alpha}{2}}   \, \,\,&\,  -\bar{\sigma}\,\exp{i\frac{\alpha}{2}} 
  \\ 0 \,\,&\,   \exp{i\frac{\alpha}{2}} \end{array}$, with $\sigma = \sigma_1 + i\,\sigma_2,$ is: 
 \scriptsize\begin{multline}
\label{18}
\bm x = A_r\, \bm X \,A_r^* =\\ \begin{array} {|cc|}
 X^0 + X^3 - X \, \bar \sigma \, \exp{i\, \alpha}- \bar X \, \sigma \, \exp{-i\, \alpha} +\sigma \,\bar \sigma \,(X^0 - X^3)&\quad\bar X \, \exp{-i\, \alpha} -\bar \sigma \,(X^0 - X^3) 
  \\  \quad X \, \exp{i\, \alpha} - \sigma \,(X^0 - X^3)   \,\,&\,   X^0 - X^3\end{array}.
\end{multline}
\normalsize 

This also reads: \scriptsize{
 \begin{equation}
\label{19}
x^0 + x^3= X^0 + X^3 - \sigma \, \bar X \, \exp{-i \, \alpha}- \bar \sigma \, {X} \, \exp{i \, \alpha} +\sigma \, \bar \sigma \,(X^0 - X^3),\quad x^0 - x^3= X^0 - X^3,
\end{equation}
 \begin{equation}
\label{20}
x =X \,\exp{i\, \alpha}-\sigma \,(X^0 - X^3),\quad \mathrm{where }\quad x =x^1 +i \,x^2,\,\, X =X^1 +i \,X^2.
\end{equation}} \normalsize
\indent Uppercase letters $X^i$ (resp. lowercase $x^i$) refer to the situation before (resp. after) the introduction of the disclination in $M^4$. We assume in the sequel that the world sheet of the $r$-disclination is the plane $\{\textbf e_0 \, \textbf e_3\}$. A point in the plane
$\{\textbf e_1 \, \textbf e_2\}$ is defined by the coordinates $x^1, \,x^2$ or alternatively $r,\, \phi$. \\ \vspace{-10pt}

When traversing a loop which surrounds the world sheet, the angle $\phi$ scans the interval $\left[0 , 2\, \pi\right]$ whereas $\Phi$ scans the interval $\left[0, 2\, \pi - \alpha\right]$. The displacement on the cut hypersurface is the sum of a rotation by an angle $\alpha$ and of a translation of components $\sigma_1, \sigma_2$. Writing that the increment of $\Phi$ is  proportional to $\phi$ when the loop is traversed, i.e., $\Delta \phi = \phi - \Phi = {\alpha \, \phi}/(2\, \pi)$, and that the translational component is also proportional to $ \frac{\phi}{2\, \pi}$, hence denoting
\small{$$U=X^0- X^3, \,V = X^0+X^3,\,u=x^0- x^3, \,v = x^0+x^3,  \, x=x^1 +i\, x^2, \, w =  \frac{u \, \phi}{2\, \pi} =  \frac{U \, \phi}{2\, \pi},$$}
\normalsize
the Eq.~\ref{19} and \ref{20} yield:
 \begin{equation} \label{21}
v-V =- \sigma \, \bar{X} \, \exp{-i \, \frac{\alpha \, \phi}{2\, \pi}}- \bar \sigma \, {X} \, \exp{i \, \frac{\alpha \, \phi}{2\, \pi}} +\sigma \, \bar \sigma \,U\,\frac{ \phi}{2\, \pi}, \quad u-U = 0,\end{equation}
 \begin{equation}
\label{22}
x = X \,\exp{i\, \frac{\alpha \, \phi}{2\, \pi}}-\sigma \,U\, \frac{ \phi}{2\, \pi},
\end{equation}
   The line element $dS^2 = -dU\,dV + dX_1^2 +dX^2_2$ of the undeformed spacetime is transformed into the line element $ds^2 = -du\,dv + dx_1^2 +dx^2_2$ in the presence of the disclination.\\ \vspace{-10pt}
 
The calculation uses the full Eq.~\ref{21} and \ref{22}; we assume that the cut hypersurface $\Sigma_{r_{\alpha}}$ is the hypersurface
\begin{equation}
\label{23}
\Sigma_{r_{\alpha}} \equiv \big\{ \half (\sigma \, \bar{X}\, \exp{-i \, \frac{\alpha \, \phi}{2\, \pi}} + \bar \sigma \, {X} \, \exp{i \, \frac{\alpha \, \phi}{2\, \pi}} ) = 0\big\}.
\end{equation} According to the discussion in Sect.~\ref{obscut} the result does not depend on the cut hypersurface, as long as the environment has the pCP symmetry $A_r$. The calculation of the line element requires some algebra but gives a rather simple result: 
\begin{equation}
\label{24}
ds^2 = -(d\lambda^0)^2 + (\delta \mu^1)^2+(\delta \mu^2)^2 +(d\lambda^3)^2,
\end{equation} 
where  $\lambda^0$ and $\lambda^3$ are defined as follows:
\begin{equation}\label{25}
\lambda^0 =x^0 -\frac{|\sigma |^2}{2}\, w,\,\lambda^1 =x^1 +\sigma_1\, w,\,\lambda^2 =x^2 +\sigma_2\, w,\,\lambda^3 =x^3 -\frac{|\sigma |^2}{2}\, w,
\end{equation} 
and \begin{equation}
\label{26}
 \delta \mu^1 =d\lambda^1 + \frac{\alpha}{2\, \pi}\,\lambda^2\,d\phi,\qquad  \delta \mu^2 = d\lambda^2 -\frac{\alpha}{2\, \pi}\,\lambda^1\,d\phi.
\end{equation} $ \delta \mu^1 $ and $ \delta \mu^2 $ are not total differentials. \\ \vspace{-10pt}

i)- \textsl{The case $\alpha = 0, \, \sigma \neq 0 $}.

The case $\sigma = 0$ brings us back to the uninteresting case of a $m$-form. In the case $\alpha = 0$, eq.~\ref{26} yields 
\begin{equation}\label{27}
ds^2 = -(d\lambda^0)^2 + (d\lambda^1)^2+(d\lambda^2)^2 +(d\lambda^3)^2.
\end{equation}
With these variables, the vanishing of the Riemann tensor is apparent. The search for a possible world sheet or world shell cannot be made with the coordinates $\lambda^a$, which are Minkowskian pure; we rather use the set of coordinates $u, \,v,\,r, \, \phi,$  with which the line element can be written:
\begin{equation}\label{28}
 (ds)^2=  -du\,dv+ |\sigma |^2 \, du \,dw  + dr^2 + r^2 \, d\phi^2 + |\sigma|^2  \, dw^2+ 2 |\sigma|\, dw \, d\left[r\, \cos (\phi - \phi_0) \right], \end{equation} where we have used the notation
 $\sigma = |\sigma| \, \exp i \phi_0,$ i.e. $\sigma_1 \,x^1 + \sigma_2 \, x^2 = |\sigma | \, r \, \cos (\phi - \phi_0)$.

One finds that the singularities are carried by the hypersurface $\Scal_{r0} \equiv\{\Hcal_{r0}=0\}$:
\begin{equation}\label{29}
\Hcal_{r0} =  r - \varsigma \,u\, \sin(\phi - \phi_0) , \qquad \mathrm{where}  \quad \varsigma = \frac{|\sigma|}{2\,\pi}.
\end{equation}

 The world shell $\Hcal_{r0} $ is not invariant along the $r$-form. Again, this can be cured by the addition of attached dislocations. For example, the edge dislocation along the $\{\textbf e_3 \}$ axis with Burgers vector $b_x = b\, \cos \phi_0, \,b_y = b\, \sin \phi_0,$:
 \begin{equation}
\label{30}
ds^2 = -dt^2+(dx+ \cos \phi_0 \,\frac{b}{2\, \pi}\,d\phi)^2 +(dy+\sin \phi_0 \,\frac{b}{2\, \pi}\,d\phi)^2 +dz^2. 
\end{equation}  is singular on the hypersurface
\begin{equation}
\label{31}
\Hcal_{e_{b}} =  r - \frac{b}{2\, \pi} \, \sin(\phi - \phi_0).  
\end{equation} Therefore dislocation densities $\sigma =db/du$ attached to the $r$-form have the expected behavior. \\ \vspace{-10pt}

ii)- \textsl{The general case $\alpha \neq 0, \, \sigma \neq 0 $}.

The singularities are carried by the hypersurface $\Scal\equiv\{\Hcal_{r} =0\}$:
\begin{equation}
\label{32}
\Hcal_{r} =  a\,r - {\alpha} \, \varsigma \,w \, \cos(\phi - \phi_0)-
{u}\,\varsigma\,\sin(\phi - \phi_0). 
\end{equation}
    
 iii)- \textsl{Another mixed $r$-form}

Equation \ref{6} for a $m$-form and eq.~\ref{24} above both describe a disclination of angle $\alpha$ about the sheet $x^1 =x^2 = 0$ in the first case, $\lambda^1 =\lambda^2 = 0$ in the second; they coincide in the exchange $x^0 \leftrightarrow \lambda^0, x^i \leftrightarrow \lambda^i$. Consequently one can use the model of a $m$-twist disclination Eq.~\ref{8} to construct another $r$-disclination with a twist component. One gets:
\begin{equation}\label{33}
ds^2 = -(d\lambda^0)^2+(d\lambda^1+ \frac{\alpha}{2\, \pi}\,\lambda^3\,d\phi)^2 +(d\lambda^2)^2 +(d\lambda^3- \frac{\alpha}{2\, \pi}\,\lambda^1\,d\phi)^2.\end{equation}

\section*{Acknowledgments} I thank Jacques Friedel, Tom Kibble and Alex Vilenkin for discussions.

 \newpage
\footnotesize
\bibliographystyle{unsrt}
\bibliography{biblio}
\end{document}